\documentclass[footinbib,aps,prmaterials,twocolumn,superscriptaddress,groupedaddress,a4paper]{revtex4}  
\usepackage{graphicx}
\usepackage{dcolumn}
\usepackage{bm}
\usepackage{amssymb}
\usepackage{amsmath}
\usepackage{subfigure}
\usepackage{color}
\usepackage{lscape}
\usepackage{siunitx}
\usepackage{afterpage}

\graphicspath{ {./Figures/}}

\begin{document}

\widetext
\title{Compositional dependence of direct transition energies in Si$_x$Ge$_{1-x-y}$Sn$_y$ alloys lattice-matched to Ge/GaAs}


\author{Phoebe M.~Pearce}
\email{p.pearce@unsw.edu.au} 
\altaffiliation[Present address: ]{School of Photovoltaic and Renewable Energy Engineering, University of New South Wales,
Sydney, NSW, Australia} 
\affiliation{Department of Physics, Imperial College London, South Kensington, London SW7 2AZ, United Kingdom}

\author{Sheau Wei Ong}
\affiliation{$\varepsilon$MaGIC-Lab, Department of Physics, National University of Singapore, 2 Science Drive 3, Singapore 117551, Singapore}

\author{Andrew D.~Johnson}
\affiliation{IQE plc., Pascal Close, St. Mellons, Cardiff CF3 0LW, United Kingdom}

\author{Eng Soon Tok}
\affiliation{$\varepsilon$MaGIC-Lab, Department of Physics, National University of Singapore, 2 Science Drive 3, Singapore 117551, Singapore}

\author{Nicholas J.~Ekins-Daukes}
\affiliation{Department of Physics, Imperial College London, South Kensington, London SW7 2AZ, United Kingdom}
\affiliation{School of Photovoltaic and Renewable Engineering, University of New South Wales, Sydney, New South Wales 2052, Australia}

\date{\today}


\begin{abstract}
Si$_x$Ge$_{1-x-y}$Sn$_y$ ternary alloys are a candidate material system for use in solar cells and other optoelectronic devices. We report on the direct transition energies and structural properties of Ge-rich Si$_x$Ge$_{1-x-y}$Sn$_y$ alloys with six different compositions up to 10\% Si and 3\% Sn, lattice-matched to Ge or GaAs substrates. The direct transitions occurring between 0.9 and 5.0 eV were investigated using spectroscopic ellipsometry (SE), and the resulting data was used to obtain the dielectric functions of the Si$_x$Ge$_{1-x-y}$Sn$_y$n layer by fitting a multi-layer model. Values for the $E_0$, $E_1$, $\Delta_1$, $E_0'$ and $E_2$ transition energies were then found by differentiating these dielectric functions to extract the locations of critical points. Structurally, the composition of the samples was measured using energy-dispersive X-ray measurements (EDX). The lattice constants predicted from these compositions are in good agreement with reciprocal space maps obtained through X-ray diffraction (XRD). The results confirm that a 1 eV direct absorption edge can be achieved using relatively low Si and Sn fractions ($<$ 10 \% and $<$ 3 \% respectively), while the higher-energy critical points show smaller shifts relative to Ge and match results previously observed or predicted in the literature.
\end{abstract}


\maketitle


\section{Introduction}
\label{sec:introduction}

Si$_x$Ge$_{1-x-y}$Sn$_y$ ternary alloys have been a topic of research interest for a variety of applications, including group IV lasers \cite{Sun2010a, Zhou2019}, photodiodes \cite{Beeler2012}, light-emitting diodes (LEDs) \cite{Beeler2013, VondenDriesch2017} and photovoltaics \cite{Fang2008, Ventura2015, Beeler2013, Roucka2016a}. By controlling the alloy composition, it is possible to tune the bandgap, higher-energy interband transitions, and the lattice constant. For photovoltaic applications, semiconductors with an absorption edge around 1 eV are an area of current research interest for current-matched in multi-junction solar cells using well-established material systems such as Ge, GaAs and In$_{0.5}$Ga$_{0.5}$P. However, identifying materials which can be grown with sufficiently high material quality at the required bandgap and lattice constant is a challenge. Si$_x$Ge$_{1-x-y}$Sn$_y$ can be engineered to have the same lattice constant as Ge but a higher bandgap and is therefore a suitable candidate material for these applications. 

Si$_x$Ge$_{1-x-y}$Sn$_y$ alloys with a direct absorption edge around 1 eV have a high atomic fraction of Ge ($x >$ 85 \%) and an atomic Si:Sn ratio of around 3.7:1, necessary to maintain the desired lattice constant according to Vegard's law \cite{Vegard1921}. The addition of Si to Ge increases the energy of both the fundamental bandgap (0.67 eV in pure Ge \cite{Adachi1988}) and the lowest-energy direction transition (0.80 eV in pure Ge \cite{Adachi1988}) but will reduce the lattice constant as Si has a smaller atomic radius than Ge. Conversely, the addition of Sn increases the lattice constant and reduces the fundamental bandgap. By balancing the relative amounts of Si and Sn in the alloy, it is possible to engineer a material which has a higher bandgap and absorption edge than Ge but the same lattice constant, although this material is expected to retain an indirect bandgap lower in energy than the 1 eV transition; it is not possible to achieve a 1 eV direct fundamental bandgap \cite{Moontragoon2012, Pearce2022}. We have previously reported on the near-bandgap behaviour of Si$_x$Ge$_{1-x-y}$Sn$_y$ alloys, investigating both the indirect fundamental bandgap and the circa-1 eV direct transition through spectroscopic measurements and theoretical calculations \cite{Pearce2022}. Here, we confirm the circa-1 eV direct absorption edge and investigate the higher-energy interband transitions for the same set of samples.

High-quality Si$_x$Ge$_{1-x-y}$Sn$_y$ alloys have been grown mainly, and most successfully, through chemical vapour deposition (CVD) and molecular beam epitaxy (MBE) \cite{Wirths2016}. Ultra-high vacuum CVD (UHV-CVD) growth of Si$_x$Ge$_{1-x-y}$Sn$_y$ was first reported by Bauer et al. \cite{Bauer2003}, followed by the first reports of photoluminescence signal from Si$_x$Ge$_{1-x-y}$Sn$_y$ by Soref et al. \cite{Soref2007}. These samples were grown using deuterated tin (SnD$_4$); this precursor was also used in the fabrication of the most successful demonstration to date of a multi-junction cell with a Si$_x$Ge$_{1-x-y}$Sn$_y$ sub-cell \cite{Roucka2016a}. Si$_x$Ge$_{1-x-y}$Sn$_y$ has also been grown through both reduced-pressure CVD, using the commercially available precursor SnCl$_4$, \cite{Wirths2013, Wirths2014, Wirths2014a} and MBE \cite{Wendav2016, Fischer2017, Lin2012}. This development of Si$_x$Ge$_{1-x-y}$Sn$_y$ epitaxy has happened in conjunction with the epitaxy of direct-bandgap GeSn alloys for laser applications, mostly in heterostructures containing (Si)GeSn. For research applications, custom-built epitaxy reactors or highly-customized commercial reactors are often used, in combination with different precursors for Si and Ge depending on the composition being grown \cite{DCosta2010}. In addition to SnD$_4$, silicon and more complex germanium hydrides (Ge$_n$H$_{2n+2}$ and Si$_n$H$_{2n+2}$) with $n > 1$ are commonly used, which are more expensive than the common commercial precursors. The samples investigated here were grown using a commercial CVD reactor and precursor materials, with the aim of developing a Si$_x$Ge$_{1-x-y}$Sn$_y$ growth process which reduces the complexity and cost associated with multi-junction cell growth. To study the direct inter-band transitions which dominate the dielectric function of these materials, spectroscopic ellipsometry was performed over the wavelength range 250--1800 nm (0.69--4.96 eV). The composition of the samples was measured through energy-dispersive X-ray spectroscopy (EDX) using a scanning electron microscope (SEM) while structural features were studied using X-ray diffraction (XRD) and optical microscopy. These results confirm that with these growth methods, it is possible to fabricate material with a 1 eV absorption edge, significantly blueshifted from pure Ge, on Ge or GaAs substrates, providing an important step towards successfully fabricating a multi-junction cell incorporating Si$_x$Ge$_{1-x-y}$Sn$_y$.

\section{Methods}

\subsection{Sample growth}
\label{sec:samples}

\begin{figure}
 \centering
 \includegraphics[width=0.5\textwidth]{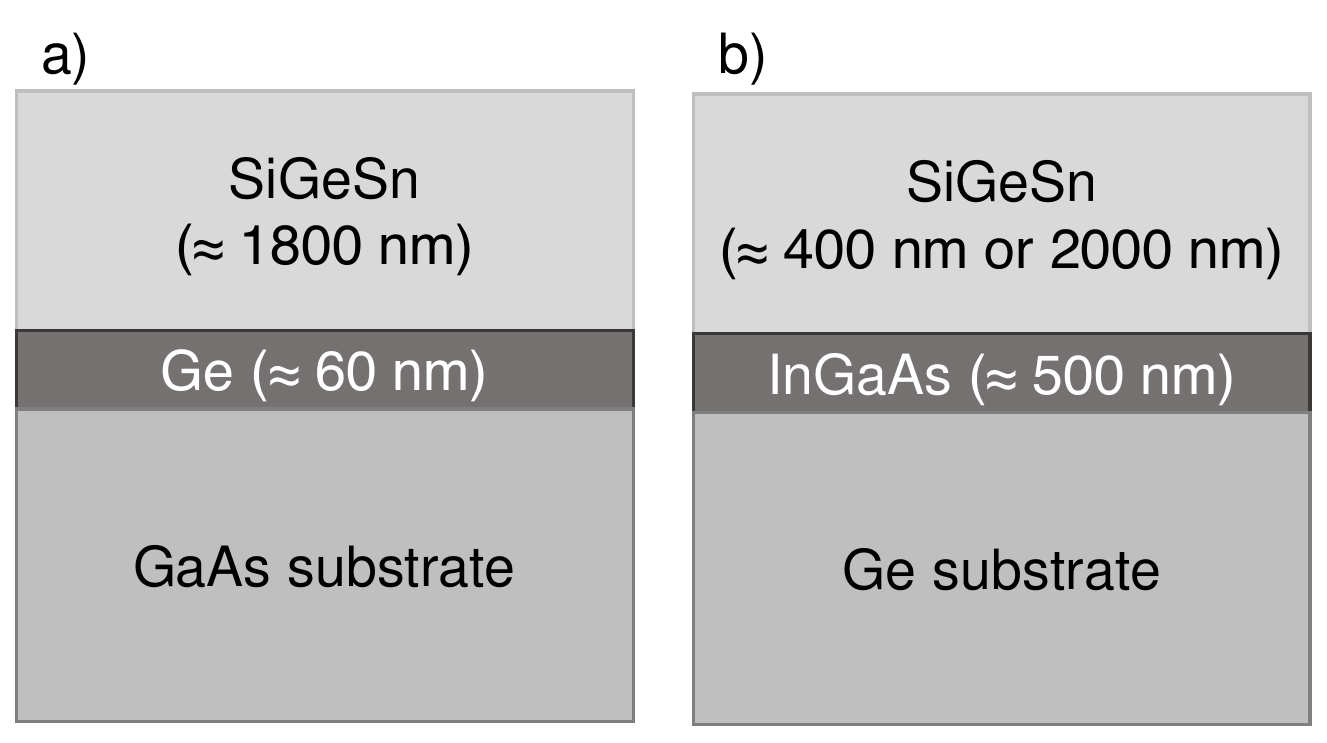}
 \caption{\label{fig:layerstructure}Schematic of the layer structure of the epitaxial Si$_x$Ge$_{1-x-y}$Sn$_y$ samples. From \cite{Pearce2022}.}
\end{figure}

Si$_x$Ge$_{1-x-y}$Sn$_y$ samples were grown through low-pressure CVD (chemical vapour deposition) in an ASM Epsilon 2000 CVD epitaxy reactor. Germane (GeH$_4$), disilane (Si$_2$H$_6$) and tin chloride (SnCl$_4$) precursors were used, with H$_2$ as the carrier gas. Two sets of samples across the composition range are considered here: one set grown on GaAs substrates with a thin ($\approx 60$ nm) Ge seed layer, which will be referred to as Set A, and one set grown on Ge substrates which were overgrown with approximately 500 nm of lattice-matched In$_{0.012}$Ga$_{0.988}$As grown through metal-organic vapour-phase epitaxy (MOVPE), referred to as Set B. The layer structure of the samples is shown in Fig. \ref{fig:layerstructure}. Samples within both of these sets were grown at three different compositions, all aiming for an approximately 3.7:1 ratio of Si:Sn to achieve the same lattice constant as Ge, assuming the lattice constant of the alloy obeys Vegard's law \cite{Vegard1921, Madelung2002}. For Set B, two different Si$_x$Ge$_{1-x-y}$Sn$_y$ thicknesses were grown for each composition, giving a total of three samples in Set A and six samples in Set B. The resulting sample structures, compositions, layer thicknesses and the labels used for the samples are summarized in Table \ref{tab:samples}.

\begin{table*}[]
\caption{\label{tab:samples}Si$_x$Ge$_{1-x-y}$Sn$_y$ thickness (from fits to SE data) $d_{SiGeSn}$, composition of the Si$_x$Ge$_{1-x-y}$Sn$_y$ measured through SEM-EDX, and resulting lattice constant obtained via Vegard's law. The samples are labelled numerically within sets in order of increasing Si/Sn composition. The composition values are the mean across ten measurements of each sample (the values in brackets are the standard deviation). The relaxed Si$_x$Ge$_{1-x-y}$Sn$_y$ lattice constant $a_{SiGeSn}$ is calculated according to Vegard's law \cite{Vegard1921} based on the compositions (the uncertainty is due to the error carried forward from the composition measurements).}
\begin{tabular}{lllll}
\hline
\textbf{Sample} & $\mathbf{d_{SiGeSn}}$ \textbf{(nm)}  & \textbf{Si (atomic \%)} &  \textbf{Sn (atomic \%)} & $\mathbf{a_{SiGeSn}}$ \textbf{(Å)} \\ \hline
A1   & 1768     &  4.6 (0.2) & 0.9 (0.1)       & 5.65(4)   \\
A2   & 1851    & 6.8 (0.5) & 1.6 (0.1)  & 5.65(5) \\
A3   & 1906   &   9.6 (0.4) & 2.2 (0.1)      & 5.65(4)   \\
B1 (thin)       & 335    & n.m. & n.m.   & 5.65(8)     \\
B2 (thin)       & 367    & n.m.  & n.m.    & 5.65(7)    \\
B3 (thin)       & 365   &  n.m.   & n.m.    & 5.65(7) \\
B1 (thick)      & 2036  & 2.6 (0.3) & 0.8 (0.2)          & 5.65(8)\\
B2 (thick)      & 2115   &   5.9 (0.2) & 1.5 (0.1)    & 5.65(7) \\
B3 (thick)      & 2080   &    8.4 (0.2)	& 2.5 (0.1) &  5.65(7)
\end{tabular}
\end{table*}

\subsection{Structural characterization}

The samples' surface morphology was examined by taking optical images using an Olympus BX51 Microscope equipped with polarizer, rotatable analyzer and Nomarski prism taken at 100$\times$ magnification.  Compositions were obtained from energy dispersive X-ray measurements (EDX) performed using a JEOL JSM-6700F Field-emission Scanning Electron microscope equipped with Oxford instruments X-Max 150 mm$^2$ taken at an accelerating voltage of 15 kV. The sample composition was measured at 10 points on the surface of each sample (averages and standard deviations are reported in Table \ref{tab:samples}); measurements of the same samples during different measurement sessions showed these results to be repeatable. The crystalline quality and lattice constant of the samples was investigated  using a Malvern-Panalytical Empyrean X-ray diffractometer by measuring the symmetrical (004) and asymmetrical (224) rocking curves and reciprocal space maps (RSMs).

\subsection{Spectroscopic ellipsometry}

Spectroscopic ellipsometry (SE) measurements were performed using a J. A. Woollam Co. VASE ellipsometer over a spectral range of 250-1800 nm at angles 75, 77 and 79$^\circ$. Ellipsometry measures the ratio of the complex reflection amplitudes $r_s$ and $r_p$ for incident polarized light, and the data is usually expressed in terms of the angles $\Psi$ and $\Delta$:

\begin{equation}\rho=\frac{r_{p}}{r_{s}}=\tan \Psi \cdot e^{i \Delta}\end{equation}

$\Psi$ quantifies the magnitude of the ratio and lies in the range $0-90^\circ$, while $\Delta$ determines the phase and lies in the range $0-180^\circ$. 

\begin{figure*}[]
 \centering
 \includegraphics[width=0.97\textwidth]{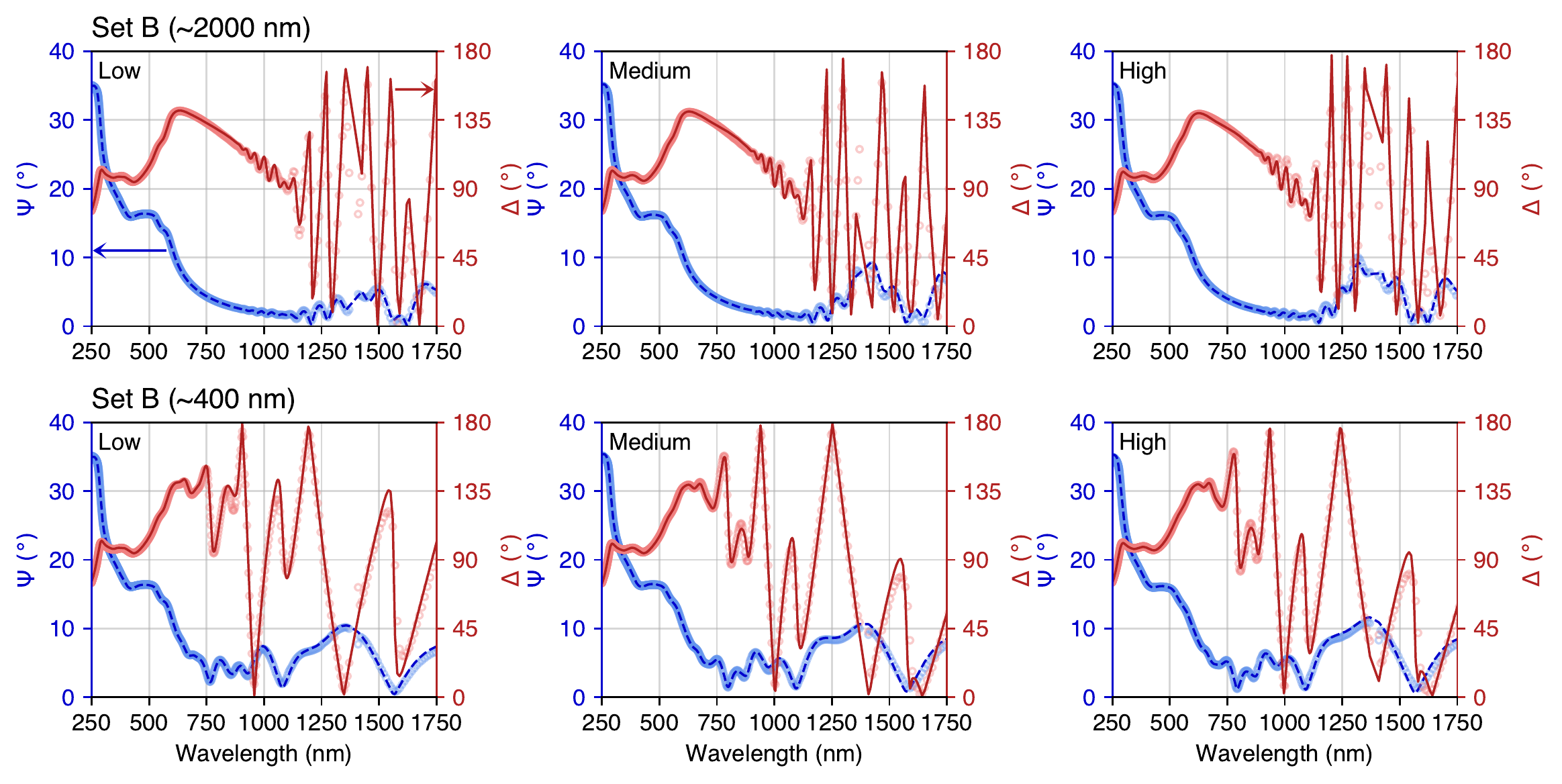}
 \caption{Ellipsometric angles $\Psi$ (left axis, in blue) and $\Delta$ (right axis, in red) for the Set B samples, as measured (open circles) with light incident at 77$^\circ$ and as calculated through the fitted multi-layer model (lines). The top row shows the results for the thicker samples in order of increasing Si/Sn composition, while the bottom row shows the results for the thinner samples. The Si$_x$Ge$_{1-x-y}$Sn$_y$ layer thicknesses fitted in each case are given in Table \ref{tab:samples}.}
 \label{fig:sigesnSE}
\end{figure*}

For a sample with multiple layers, it is generally necessary to fit a multi-layer model to the data which accounts for reflection at the front surface and each interface, and the resulting interference at each wavelength; such interference is evident in the data shown in Fig. \ref{fig:sigesnSE}. A parametric Herzinger-Johs \cite{Johs1998} model was used to fit the Si$_x$Ge$_{1-x-y}$Sn$_y$ optical constants. Such a parametric model has the advantages of reducing the number of fitting parameters (compared to fitting the $\varepsilon_1$ and $\varepsilon_2$ values point-by-point) and enforcing Kramers-Kronig consistency \cite{J.A.WoollamCo.2012, Johs1998}, while increasing the flexibility of the peak shapes compared to a more physically-constrained model such as a Critical Point Parabolic Band (CPPB) model \cite{Adachi2005}. The drawback compared to a CPPB model is that due to the flexible peak shape, the fitting parameters drawn from a parametric model are not necessarily physically meaningful; e.g. the centre energy of a peak may not correspond directly to a specific transition energy in the material's band structure. For this reason, rather than taking the desired transition energies directly from the obtained fitting parameters, the dielectric function obtained for the Si$_x$Ge$_{1-x-y}$Sn$_y$ was differentiated in order to obtain the relevant centre energies, as described below.

The structure of the different sample types, used as a starting point for these models, is shown in Fig. \ref{fig:layerstructure}, comprising the Ge or GaAs substrate, the InGaAs layer (in the case of the Set B samples), and the Si$_x$Ge$_{1-x-y}$Sn$_y$ layer. A GeO$_2$ surface layer \cite{Fleming1984} was also included in the fits; the inclusion of a thin surface layer (fitted as 1-2 nm thick for all the Si$_x$Ge$_{1-x-y}$Sn$_y$ samples) is necessary to achieve a good fit to the data, including for a reference measurement of a Ge substrate. GeO$_2$ was used since the samples are mostly Ge, so it was assumed the surface oxide is mostly Ge-based. A similar method to the fitting procedure for the Si$_x$Ge$_{1-x-y}$Sn$_y$ optical constants outlined in \cite{DCosta2006b} was used. The starting points for the layer thicknesses were the nominal thicknesses targeted during CVD growth. Initially, the Si$_x$Ge$_{1-x-y}$Sn$_y$ layer was assumed to be pure Ge; since we are considering Ge-heavy ($>$ 85\%) samples of Si$_x$Ge$_{1-x-y}$Sn$_y$, using the well-known optical constants of Ge as a placeholder for Si$_x$Ge$_{1-x-y}$Sn$_y$ gives a good starting point for fitting the layer thicknesses. For the Set B samples, the InGaAs buffer layer thickness was also allowed to vary at this stage; in each case the deviation from the nominal thickness of 500 nm was found to be $\leq$ 10 nm. The starting point for the Herzinger-Johs model fits for Si$_x$Ge$_{1-x-y}$Sn$_y$ was the built-in model for Ge \cite{J.A.WoollamCo.2012}. The model allows for very flexible peak shapes; each feature is described by seven variable parameters which control the peak's strength, centre energy, and broadening, with further parameters controlling the shape and asymmetry of the peak. Here, only the peak strength ($A$), centre energy ($E_0$) and broadening ($B$) were fitted, with the other parameters kept fixed at the Ge values. This was to avoid unreasonable peak shapes, including some with discontinuities in the dielectric function, especially in the low-energy region where thin-film interference and noise due to signal depolarization affects the SE measurements\footnote{Depolarization occurs when the bulk of the sample becomes transparent, and thus the incident polarized light is able to travel through the sample, reflecting at the back surface and returning to the front of the sample, where it can be collected by the detector. As this light has travelled a comparatively large distance through the sample (hundreds of microns compared to hundreds of nanometres), it is no longer coherent with the rest of the signal and causes depolarization of the measured signal.}. It was found that excellent fits to the data could be achieved without varying the other peak shape parameters. The parameters for the Si$_x$Ge$_{1-x-y}$Sn$_y$ optical constants were fitted starting at high energies, initially fitting only the centre energies of the peak followed by allowing the peak strength and broadening to vary. The layer thickness of the Si$_x$Ge$_{1-x-y}$Sn$_y$ is allowed to continue to vary while the optical constants are fitted. The data measured at all the incidence angles (75, 77 and 79$^\circ$) are fitted simultaneously. The optical constants for the other layers (InGaAs, Ge and GaAs) came from the WVASE software database, except for those of GeO$_2$ which were defined according to a Sellmeier model using parameters from \cite{Fleming1984}.

After the layer thicknesses have been determined exactly, and a parametric model for the Si$_x$Ge$_{1-x-y}$Sn$_y$ optical constants has been fitted, a point-by-point fit where $\varepsilon_1$ and $\varepsilon_2$ are fitted separately at each measurement wavelength is also performed, keeping the layer thicknesses fixed. The two fitting methods agree very well for energies above $\approx$ 1 eV (depending on the Si$_x$Ge$_{1-x-y}$Sn$_y$ composition). Below these energies (i.e. below the lowest direct transition in the Si$_x$Ge$_{1-x-y}$Sn$_y$, where Si$_x$Ge$_{1-x-y}$Sn$_y$ is only very weakly absorbing) the presence of thin-film interference fringes in the data significantly affects the point-by-point fits, as shown in the Supplemental Information \cite{supmat}. 

The fundamental indirect gap -- 0.67 eV for Ge at room temperature and blueshifted to 0.7--0.8 eV for the Si$_x$Ge$_{1-x-y}$Sn$_y$ samples, as determined by photoluminescence measurements \cite{Pearce2022}) -- is not included in the Herzinger-Johs model. Even considering the expected blueshift in this bandgap relative to Ge for the Si$_x$Ge$_{1-x-y}$Sn$_y$ samples, this transition occurs very close to or beyond the longest wavelength which can be measured using the V-VASE ellipsometer, and is expected to contribute so weakly to the dielectric function that it would be extremely difficult to observe through ellipsometry. This transition was investigated through photoluminescence measurements of the same set of samples as reported in \cite{Pearce2022}. 

\subsection{Critical point fits}

The optical constants obtained from the SE data were differentiated twice to obtain fits for the centre energies of the peaks identified as corresponding to the $E_0$, $E_1$, $E_1 + \Delta_1$, $E_0'$ and $E_2$ transition energies, using a method similar to \cite{DCosta2006a}. The critical points observed in the second derivative of the dielectric function were fitted to expressions of the form:

\begin{equation}\label{eq:CP}
\frac{d^2(E^2\varepsilon)}{dE^2}=\frac{A_ie^{i\Phi_i}}{(E-E_i+i\Gamma_i)^m}
\end{equation}

where the strength parameter $A_i$, phase $\Phi_i$, centre energy $E_i$ and broadening term $\Gamma_i$ are fitting parameters. The exponent $m$ is fixed and depends on the type of critical point (CP) being fitted. The $E_0$ and $E_0 + \Delta_0$ features were fitted as a 3D M0 type CPs with the exponent $m = 3/2$, while the higher-energy CPs were fitted as 2D M1 type CPs with $m = 2$. The choice of this differential (specifically the choice to differentiate the quantity $E^2\varepsilon$ rather than the more common choice of simply differentiating $\varepsilon$), and how the differentiation was performed computationally, are discussed in the Supplemental Material \cite{supmat}.

As described above, the point-by-point fit was reliable above $\approx$ 1 eV, while below these energies it was very noisy due to the interference fringes in the measurement and signal depolarization. Thus, for the analysis of ellipsometric data presented in Section \ref{sec:CPfits}, the point-by-point fits were used for energies $>$ 1.5 eV to fit the $E_1$, $E_1 + \Delta_1$, $E_0'$ and $E_2$ centre energies while the Herzinger-Johs parametric model fit was used to obtain $E_0$. 


\begin{figure}[]
 \centering
 \includegraphics[width=0.5\textwidth]{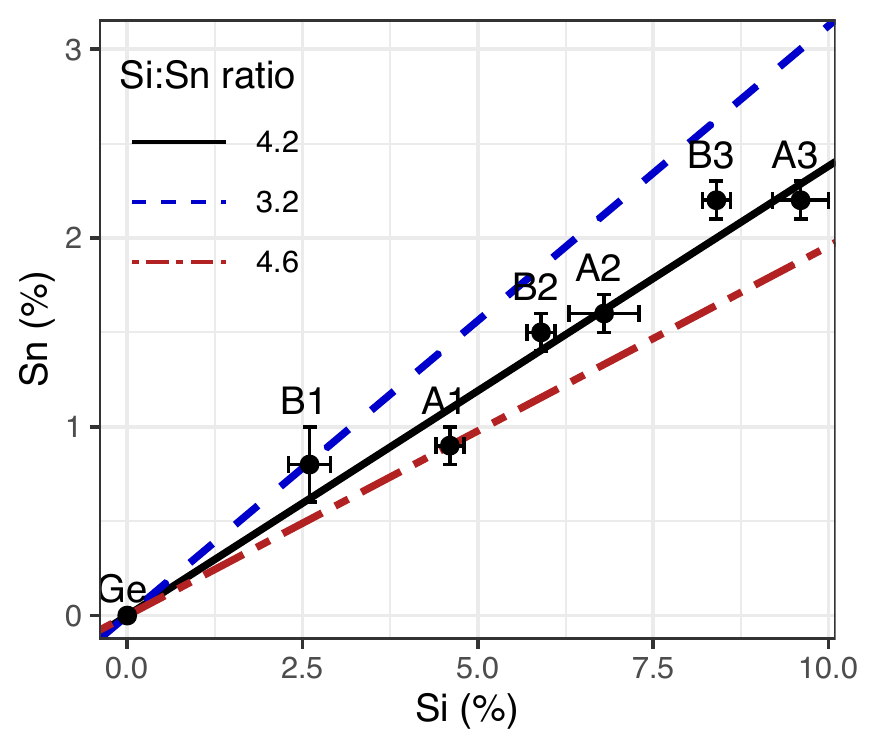}
 \caption{\label{fig:Si_Sncomp}The Si:Sn ratio for each of the samples with composition measured through SEM-EDX. For Set B, only the thicker samples were measured. The error bars show the standard deviation from ten measurements of the same sample.}
\end{figure}

\begin{figure*}[]
 \centering
 \includegraphics[width=0.9\textwidth]{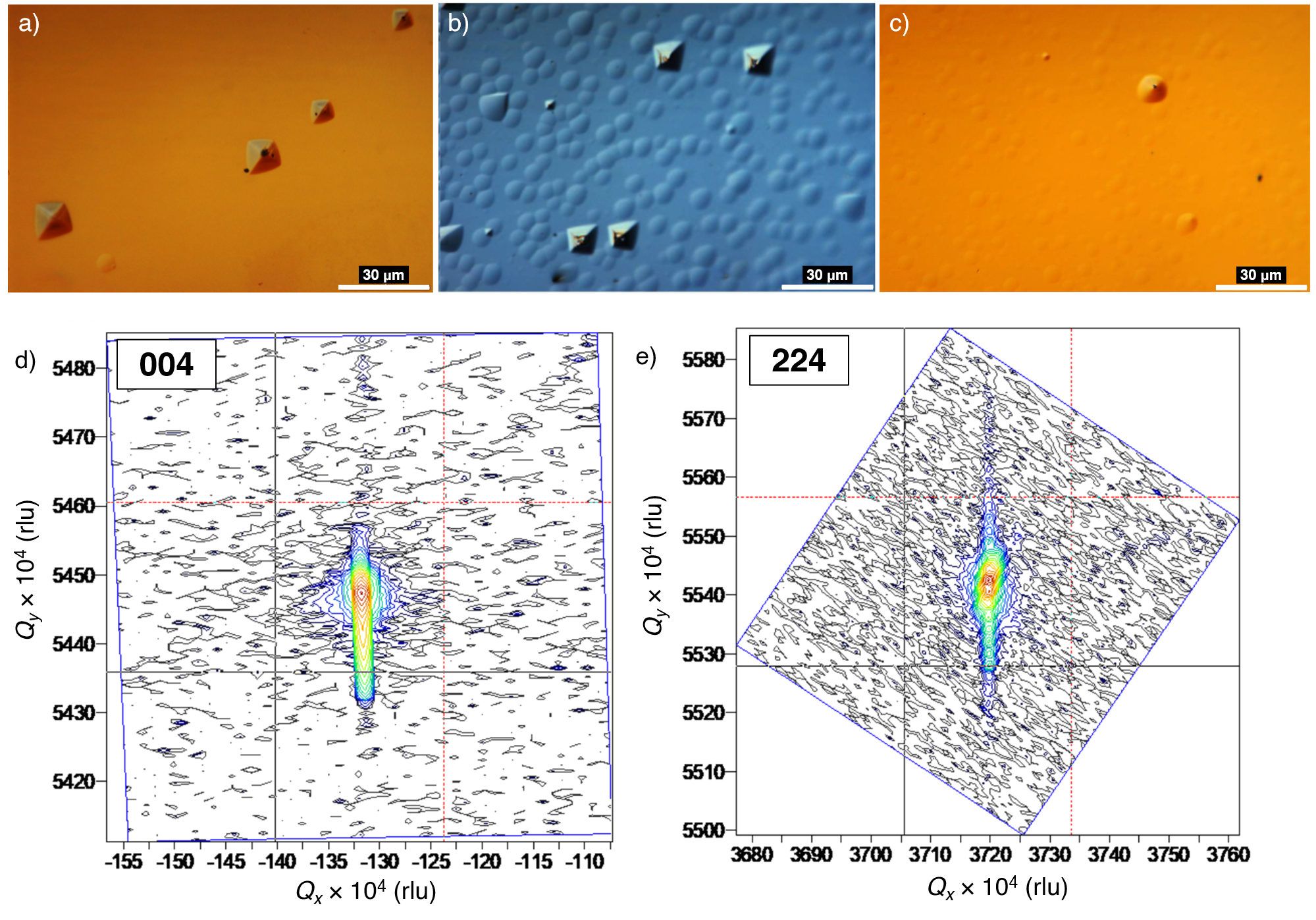}
 \caption{\label{fig:sigesnstructural}
 (a)-(c) Optical microscopy image of the surface of samples A1, A2 and A3 respectively, taken at 100$\times$ magnification. (d) RSM measured around the (004) direction for sample A1. (e) RSM measured around the (224) direction for sample A1.}
\end{figure*}

\begin{figure}[]
 \centering
 \includegraphics[width=0.5\textwidth]{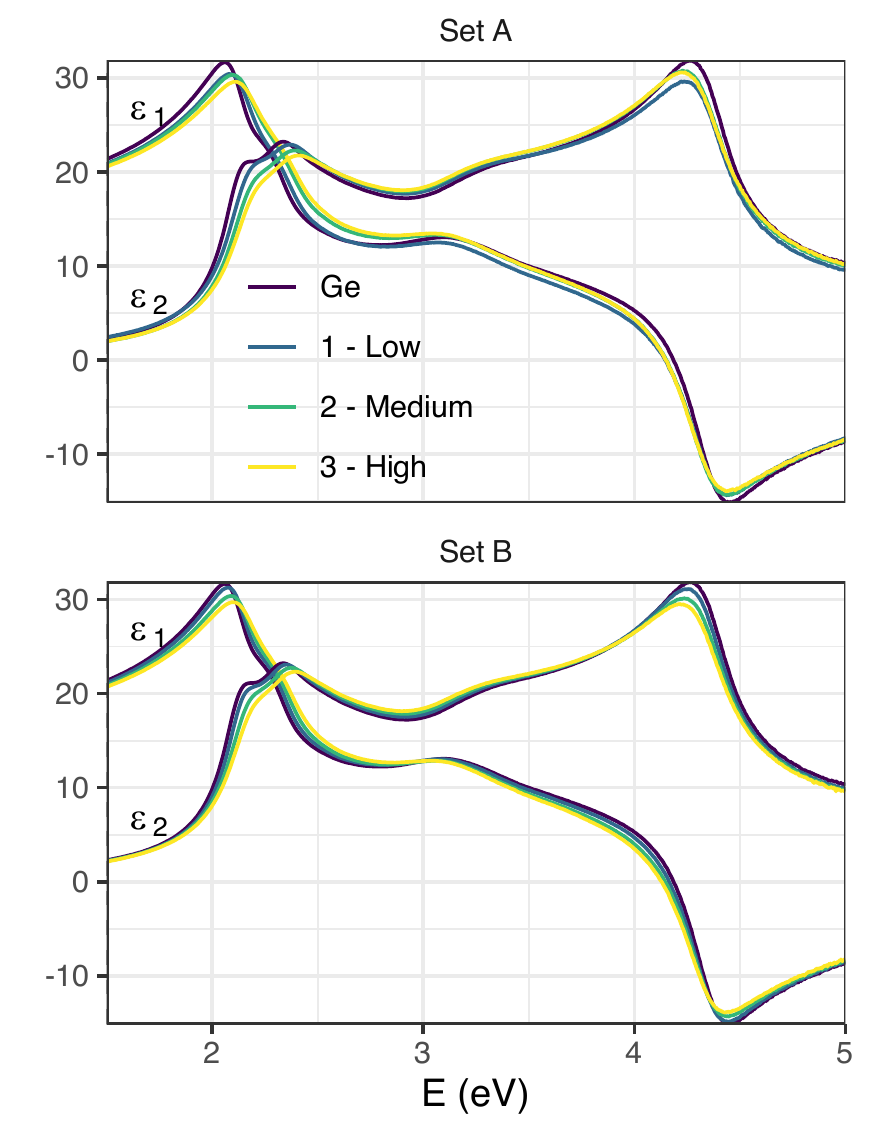}
 \caption{Real ($\varepsilon_1$) and imaginary ($\varepsilon_2$) parts of the dielectric function of the Si$_x$Ge$_{1-x-y}$Sn$_y$ samples, obtained through fitting the ellipsometry data.\label{fig:sigesn_epsilon}}
\end{figure}

\begin{figure*}[]
 \centering
 \includegraphics[width=\textwidth]{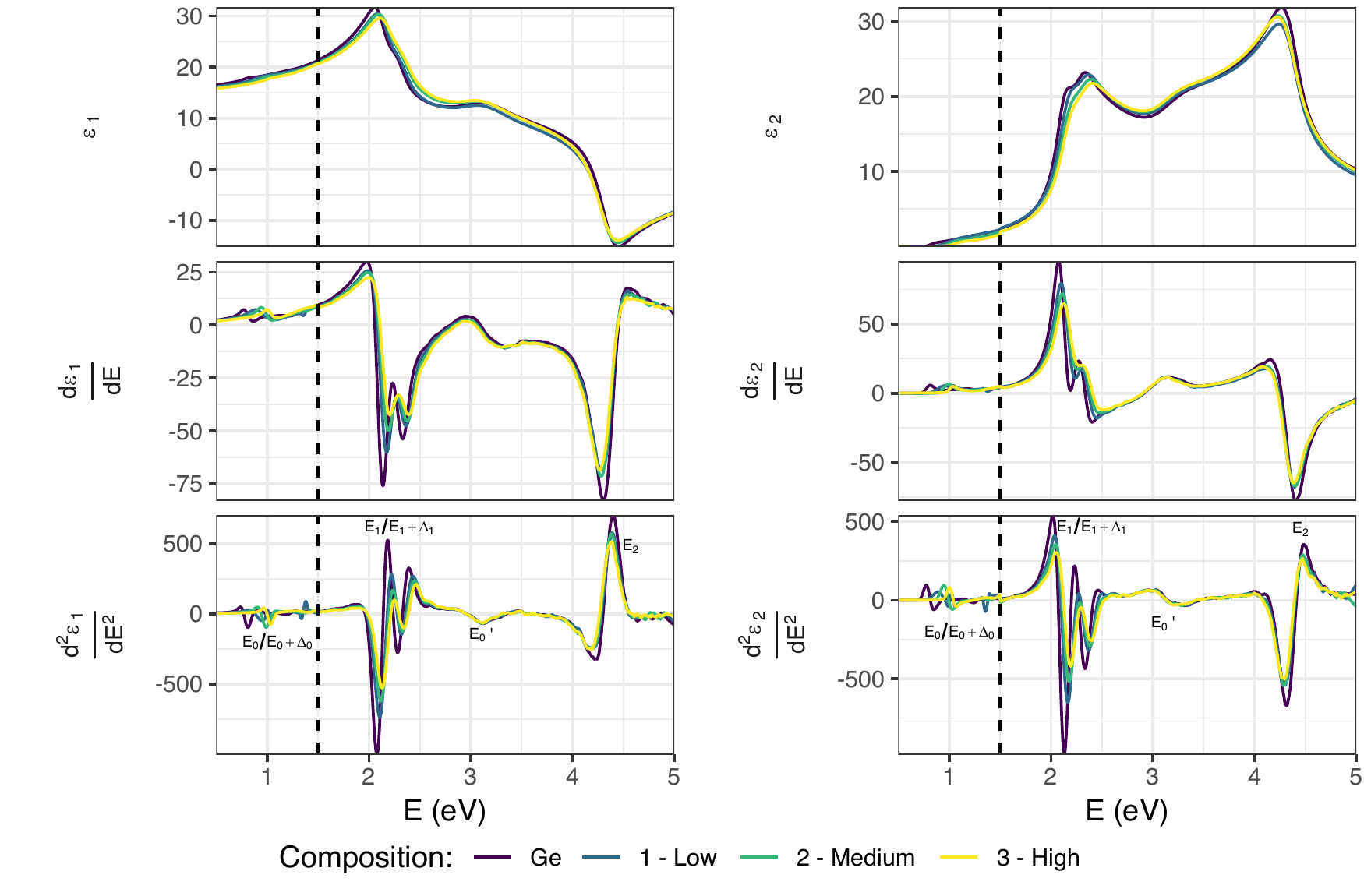}
 \caption{\label{fig:epsilon_derivs}Real ($\varepsilon_{1}$) and imaginary ($\varepsilon_{2}$) parts of the dielectric functions from fits to ellipsometry data for sample set A, and their first and second derivatives with respect to energy. Data for a Ge substrate is shown for reference in each plot. Data below 1.5 eV is from the parametric model fit, while data above 1.5 eV is from a point-by-point fit to the data; the dashed line indicates where the data is stitched together.}
\end{figure*}

\begin{figure}[]
 \centering
 \includegraphics[width=0.46\textwidth]{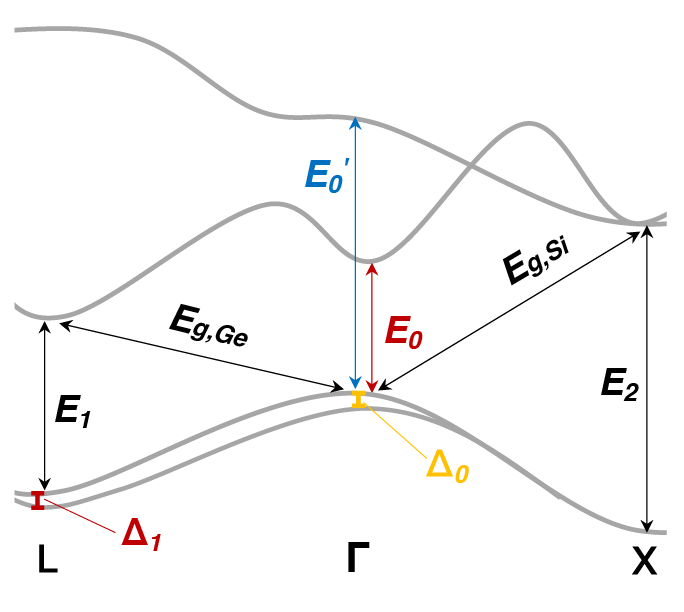}
 \caption{\label{fig:sigesnsimple}Simplified band structure of Si$_x$Ge$_{1-x-y}$Sn$_y$ showing key indirect and direct transition energies in the band structure.}
\end{figure}

\section{Results}
\label{sec:results}

\subsection{Structural measurements}

The composition and lattice constant of the Si$_x$Ge$_{1-x-y}$Sn$_y$ samples were characterized using SEM-EDX and XRD (rocking curve and reciprocal space mapping) measurements, clearly showing the increasing Si and Sn compositions within the sample sets. Table \ref{tab:samples} gives the composition obtained through EDX measurements, and the predicted lattice constant through Vegard's law. Fig \ref{fig:Si_Sncomp} shows the Si and Sn fractions of each sample; all samples lie between a 4.6:1 and 3.2:1 Si:Sn ratio, with the Set B samples having a lower ratio than the Set A samples, meaning the Set B samples are more closely lattice-matched to their Ge substrates, while the Set A samples are more closely lattice-matched to their GaAs substrates. For all the samples, the thickness of the epitaxial Si$_x$Ge$_{1-x-y}$Sn$_y$ layer is expected to be below the critical thickness, as shown in the Supplemental Material \cite{supmat}. 

Surface imaging shows the presence of defects, but no cross-hatching or features which indicate relaxation of the epitaxial layer. For Set B, only the thicker samples ($\approx 2000$ nm) were measured in these structural measurements, and it was assumed that the corresponding thinner ($\approx 400$ nm) Si$_x$Ge$_{1-x-y}$Sn$_y$ layers have very similar compositions; this is corroborated by the consistency of the dielectric function across the different thicknesses, as discussed in the next section. The surfaces of both sets of samples show `bubble' and `pyramid'- like defects, as shown in Fig. \ref{fig:sigesnstructural}(a)-(c). Given the known difficulties of Si$_x$Ge$_{1-x-y}$Sn$_y$ growth, especially regarding Sn incorporation, it is reasonable to assume these defects may be related to segregation of Sn during growth; however, additional EDX measurements comparing flat areas of the surface with the defect features do not indicate a measurably different composition within either of these feature types. The pyramid-type defects penetrate into the bulk of the material significantly, and appear to be the results of growth along a different crystal axis to the bulk film, although further investigation is required.

\begin{figure*}
\centering
 \centering
 \includegraphics[width=\textwidth]{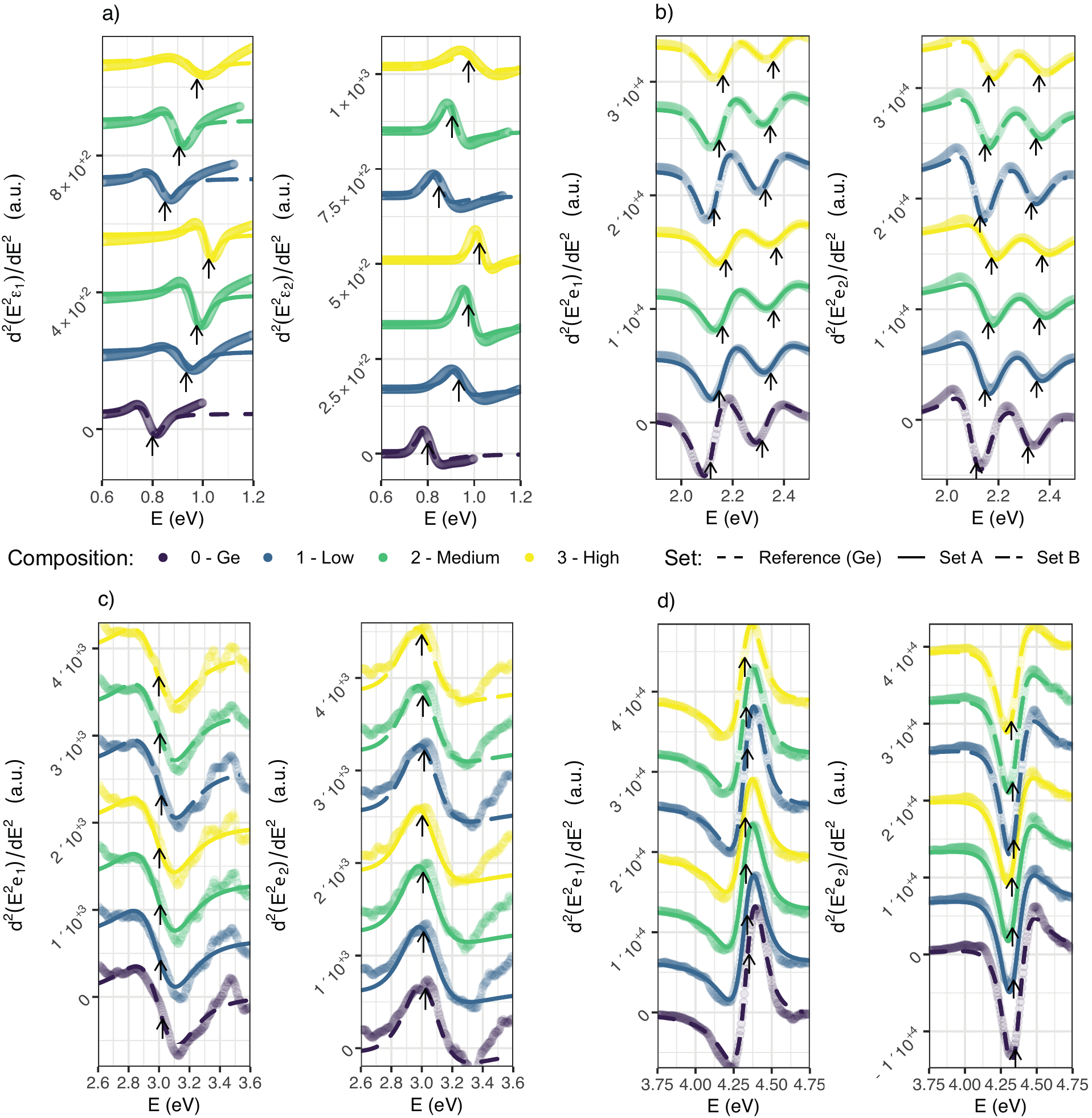}
\caption{Fits (dashed lines) of the critical point model (eq. \ref{eq:CP}) to features in the differentiated SE data (open circles) corresponding to (a) the $E_0$ transition, (b) the $E_1$ and $E_1 + \Delta_1$ transitions, (c) the $E_0'$ transition and (d) the $E_2$ transition. The arrows indicate the values of the fitted centre energies.}
\label{fig:SEderivfits}
\end{figure*}

Rocking curves and reciprocal space maps (RSMs) were measured around the (004) and (224) lattice points. Fig. \ref{fig:sigesnstructural}(d) and (e) show the RSM scans for sample A1; RSM data for the other Set A samples around both points is given in the Supplemental Material \cite{supmat}. Only the out-of-plane lattice constant can be observed in the (004) scans, while the asymmetric (224) scan will contain contributions from both the in-plane and out-of-plane lattice constants. For samples A1 and A3, some spread around a single RSM peak is observed in both the (004) and (224) RSMs, but only a single clearly-defined peak is visible, indicating that the epi-layer is very closely lattice-matched to the substrate. The peak broadening in reciprocal space at constant $Q_x$ also indicates pseudomorphic (constant lattice constant) growth, with the in-plane lattice parameter strained to the substrate (GaAs) lattice constant. For sample A2, a clear split in the peak in the (004) RSM is observed, showing that the epi-layer has a different out-of-plane lattice constant than the substrate. This indicates the presence of biaxial strain and thus that the epitaxial layer has not relaxed. The XRD peaks for this sample correspond to lattice constants of 5.6537 Å and 5.6567 Å, with the former value being the GaAs lattice constant. This indicates that the relaxed lattice constant of sample A2 is larger than that of GaAs, causing in-plane compressive strain in the Si$_x$Ge$_{1-x-y}$Sn$_y$ and expansion of the out-of-plane lattice constant. Assuming the in-plane lattice constant is pinned to that of GaAs, as indicated by the RSM in the (224) direction, and taking the Poisson ratio \footnote{The Poisson ratio describes the out-of-plane distortion for in-plane strain.} of the Si$_x$Ge$_{1-x-y}$Sn$_y$ to be 0.3 (a typical value used for Ge \cite{People1985}), the relaxed lattice constant is calculated to be 5.6553 Å, in excellent agreement with the value of 5.655 Å calculated through Vegard's law using the composition measurements from SEM-EDX (see Table \ref{tab:samples}). Sample A2 was the Set A sample with the largest predicted deviation from the GaAs lattice constant through Vegard's law, and the only sample for which a clear second peak in the RSM can be observed. This indicates that the samples are closely lattice-matched to the substrates, with some strain, and the SEM-EDX composition measurements and XRD give consistent results.

\subsection{Ellipsometry}
\label{sec:results_experiment}

The ellipsometry data for the Set B samples, and the result of the model fit, are shown in Fig. \ref{fig:sigesnSE} (data for the Set A samples is shown in the Supplemental Material \cite{supmat}, and the SE data and resulting fits of the Si$_x$Ge$_{1-x-y}$Sn$_y$ optical constants are available \cite{data}). This data shows clear effects due to thin-film interference in the Si$_x$Ge$_{1-x-y}$Sn$_y$ layer; while the higher-energy transitions (2 eV and above) are clearly visible in the raw data, the near-bandgap data is dominated by interference fringes which depend on the thickness of the Si$_x$Ge$_{1-x-y}$Sn$_y$ layer. This can be seen clearly by comparing the data in Fig. \ref{fig:sigesnSE} from the samples with the same composition but different thicknesses, with the thinner samples showing much wider interference fringes. 

For the Set B samples, the optical constants of the Si$_x$Ge$_{1-x-y}$Sn$_y$ were obtained only through fitting to the data from the thicker samples. To fit the SE data of the thinner samples, only the layer thicknesses were allowed to vary. This allowed for a validation of the optical constants fitted and a check that the Si$_x$Ge$_{1-x-y}$Sn$_y$ layer has the same composition for the different thicknesses; the models for the thinner Set B samples show excellent agreement with the data despite only the layer thicknesses being fitted as shown in Fig. \ref{fig:sigesnSE}. Fig. \ref{fig:sigesn_epsilon} shows the dielectric function fitted for each Si$_x$Ge$_{1-x-y}$Sn$_y$ composition in Set A and Set B, showing how the absorption edge and lower-energy critical points ($<$ 2.5 eV) blueshift with increasing Si/Sn composition while the higher-energy critical points redshift. Fig. \ref{fig:epsilon_derivs} shows the real and imaginary part of the dielectric function fitted to the Si$_x$Ge$_{1-x-y}$Sn$_y$ layers of the Set A samples, and their first and second derivatives. Through fitting equation \ref{eq:CP}, the exact location of the transition energies can be extracted from these derivatives; these sharp features are clearly visible in the second derivative shown in Fig. \ref{fig:epsilon_derivs}. 

\begin{figure*}[]
 \centering
 \includegraphics[width=\textwidth]{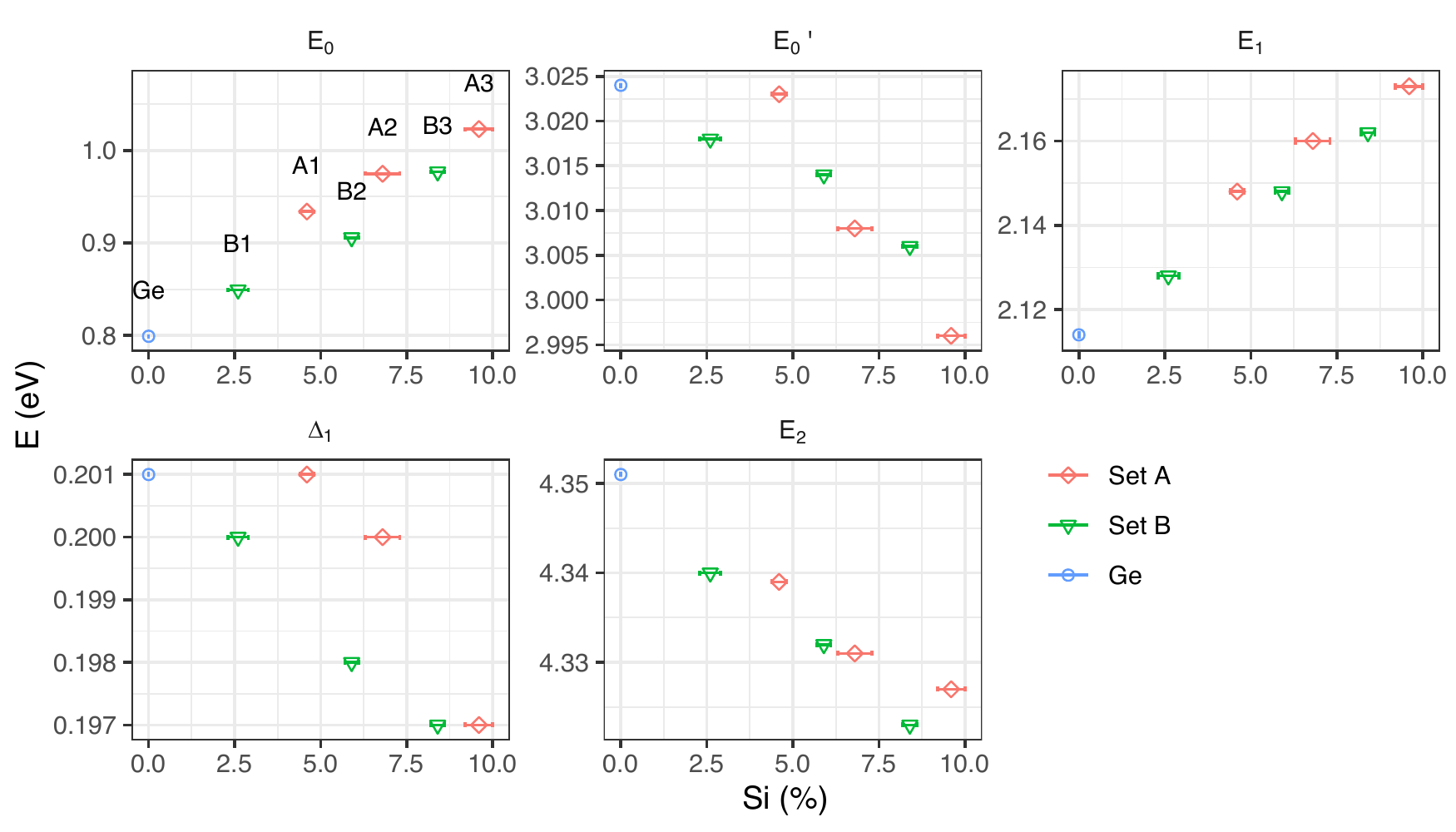}
 \caption{\label{fig:allenergies}Summary of centre energies for the different direct transitions extracted from SE data.}
\end{figure*}

\subsection{Critical point fitting}\label{sec:CPfits}

Fig. \ref{fig:sigesnsimple} shows a simplified version of the Si$_x$Ge$_{1-x-y}$Sn$_y$ band structure, indicating the point in the band structure at which the transitions discussed in this section take place. Fig. \ref{fig:SEderivfits} shows the critical point fits according to equation \ref{eq:CP} (with $m = 3/2$ for a $E_0$ and $m = 2$ for all the other features) for the $E_0$, $E_1/E_1 + \Delta_1$, $E_0'$ and $E_2$ features obtained from differentiating the dielectric functions shown in Fig. \ref{fig:sigesn_epsilon}. For each transition, the phase was fixed to be the same across compositions, choosing the phase which gives the best fit across all the samples (see Supplemental Material Table S1 for full fitting results \cite{supmat}). For $E_0$, the parametric model fits were differentiated twice, while to fit the higher-energy critical points the point-by-point dielectric constant fit was used. The results for the critical point fits are summarized in Table \ref{tab:allfits} and shown in Fig. \ref{fig:allenergies}. 

\begin{table}[]
\caption{\label{tab:allfits}Summary of centre energies fitted to the Si$_x$Ge$_{1-x-y}$Sn$_y$ and Ge critical points as obtained through SE measurements. All values are in eV.}

\begin{tabular}{llllll}
\hline
\textbf{Comp.} & $E_0$   & $E_1$  &  $E_1 + \Delta_1$   & $E_0'$   & $E_2$    \\ \hline
Ge & 0.799  & 2.114  & 2.315 & 3.024 & 4.351 \\
A1 & 0.934 & 2.148 & 2.349 & 3.011 & 4.339 \\
A2 & 0.975 & 2.160 & 2.360  & 3.008 & 4.331 \\
A3 & 1.023 & 2.173 & 2.370  & 3.002 & 4.327 \\
B1 & 0.849 & 2.128 & 2.328 & 3.016 & 4.340  \\
B2 & 0.906 & 2.148 & 2.346 & 3.007 & 4.332 \\
B3 & 0.977 & 2.162 & 2.359 & 2.999 & 4.323\\ \hline
\end{tabular}
\end{table}

As expected, the lowest direct interband transition energy $E_0$ of Si$_x$Ge$_{1-x-y}$Sn$_y$ shifts to higher energies as the Si and Sn composition increase. The values obtained here from fitting SE data are in good agreement with the results of photoreflectance measurements of the same samples reported in \cite{Pearce2022}; the deviation in the $E_0$ values between the two measurements is $<$ 4\% for all samples. A 1 eV $E_0$ transition can be achieved with $<$ 10\% Si and $<$ 3\% Sn with a Si:Sn ratio of around 4:1, while maintaining a good lattice match to Ge/In$_{0.012}$Ga$_{0.988}$As. The trend in $E_0$ shows an effect from higher relative Sn composition in the set B samples: these samples show less blueshift than the Set A samples, as can be seen in Fig. \ref{fig:allenergies}. The $E_1$ and $E_1 +\Delta_1$ energies, around 2.2 and 2.3 eV respectively, also blueshift with increased Si/Sn composition as expected \cite{DCosta2010, Xu2015}. Meanwhile, $E_0'$ and $E_2$ redshift very slightly (Fig. \ref{fig:allenergies}), each by less than 30 meV between Ge and the highest-composition sample (A3), as does the split-off energy $\Delta_1$ though by only 4 meV. The reduction in $E_0'$ matches previously reported results \cite{DCosta2010}. Although different sources report different values for the $E_0'$ energies of Ge and Si \cite{Adachi1988}, the values for Si are generally slightly higher by around 0.2 eV, yet we observe a slight redshift here; this could be due to the presence of $\alpha$-Sn (which has a lower $E_0$ by $\approx$ 1 eV) outweighing the effect of the Si, or because what is treated here as a single transition is in fact made up of at least two contributing critical points. This can be seen at the high-energy edge of Fig. \ref{fig:SEderivfits}(c), where there appears to be a second weaker feature visible around 3.4-3.5 eV. The reduction in $E_2$ is also consistent with previous literature results \cite{Aella2004,DCosta2010}, and expected from the lower $E_2$ energy of both Si and $\alpha$-Sn compared to Ge. Note that values of $\Delta_0$ were not fitted to the ellipsometry data; this contribution is relatively much weaker than that of $E_0$, which is already much weaker than the higher-energy transitions. It was not possible to obtain a reliable, unique fit to the $E_0 + \Delta_0$ critical point, so it was excluded from the second-derivative fits. Photoreflectance measurements of the same samples indicate that $\Delta_0$ stays almost constant \cite{Pearce2022}.


\section{Conclusions}
\label{sec:conclusions}

Si$_x$Ge$_{1-x-y}$Sn$_y$ is a promising candidate material for solar cells and other optoelectronic applications, with previous experimental and theoretical results indicating that a direct transition energy at 1 eV can be achieved at atomic compositions $<$ 10\% Si and $<$ 3\% Sn, with the resulting alloy being lattice-matched to Ge or GaAs. The results of ellipsometry measurements presented here confirm this, showing $E_0$ energies between 0.98 and 1.02 eV for alloys with compositions of 6.8-9.6\% Si and 1.6-2.5\% Sn. The higher-energy (up to 5 eV) direct transitions at the $\Gamma$, $X$ and $L$ points in the bandstructure were also investigated, showing a blueshift in $E_1$ and small redshifts in $\Delta_1$, $E_0'$ and $E_2$, in good agreement with previously-published results.

At these compositions, Si$_x$Ge$_{1-x-y}$Sn$_y$ retains a fundamentally indirect ``Ge-like" bandgap. This indirect transition was studied for this set of samples using photoluminescence, showing bandgaps in the range 0.7-0.75 eV \cite{Pearce2022}. While this lower-lying indirect transition is expected to affect the voltage of a solar cell adversely, Si$_x$Ge$_{1-x-y}$Sn$_y$ will have a higher fundamental bandgap and thus voltage than Ge (if used instead of Ge in e.g. a triple-junction device with InGaP and (In)GaAs). In a four-junction InGaP/(In)GaAs/SiGeSn/Ge device, where current-matching is important, absorption across the Si$_x$Ge$_{1-x-y}$Sn$_y$ will be dominated by the direct edge and the additional Si$_x$Ge$_{1-x-y}$Sn$_y$ junction will still provide a voltage boost compared to the standard triple-junction architecture. Thus, the 1 eV direct absorption edge is still expected to provide benefits in terms of current-matching, and the higher indirect gap compared to Ge can provide an increase in voltage. 


\section*{Data access}

The spectroscopic ellipsometry data associated with this work are openly available, and can be accessed via Ref.~\onlinecite{data}.

\section*{Acknowledgements}
This work was supported by the Engineering and Physical Sciences Research Council, U.K. (EPSRC; via a CASE Studentship, held by P.P. and sponsored by IQE plc.), and by the Royal Society (via an Industry Fellowship, held by N.J.E-D.).


\nocite{Matthews1974}
\nocite{Yu2010}
\nocite{Vina1984}
\nocite{Gallagher2014}
\nocite{Elzhov2016}

\bibliographystyle{apsrev} 
\bibliography     {main.bib} 

\end{document}